\documentclass[aps,prc,showpacs,floatfix]{revtex4}
\usepackage{graphicx}

\begin{document}

\draft
\title{Symmetry energy at subnuclear densities deduced from nuclear masses}

\author{Kazuhiro Oyamatsu$^{1,2}$ and Kei Iida$^{2,3}$}
\affiliation{$^1$Department of Media Theories and Production, Aichi Shukutoku
University, Nagakute, Nagakute-cho, Aichi-gun, Aichi 480-1197, Japan\\
$^2$RIKEN Nishina Center, RIKEN, Hirosawa, Wako, Saitama 351-0198, Japan\\
$^3$Department of Natural Science, Kochi University,
Akebono-cho, Kochi 780-8520, Japan
}

\date{\today}

\begin{abstract}

     We examine how nuclear masses are related to the density dependence 
of the symmetry energy.  Using a macroscopic nuclear model we 
calculate nuclear masses in a way dependent on the equation of state of
asymmetric nuclear matter.  We find by comparison with empirical two-proton
separation energies that a smaller symmetry energy at subnuclear densities,
corresponding to a larger density symmetry coefficient $L$,
is favored.  This tendency, which is clearly seen for nuclei that are
neutron-rich, nondeformed, and light, can be understood from the property of 
the surface symmetry energy in a compressible liquid-drop picture.

\end{abstract}
\maketitle

\section{Introduction}

    Saturation of the density and binding energy is a fundamental 
property of atomic nuclei.  Due to this property, the nuclear masses can be
described well using a liquid-drop approach.  Conventionally, in this approach,
the nuclear binding energy $E_B$ is written as function of mass number
$A$ and charge number $Z$ (or neutron number $N$) with the 
Weizs{\" a}cker-Bethe mass formula \cite{BW}
\begin{equation}
    -E_B=E_{\rm vol}+E_{\rm sym}+E_{\rm surf}+E_{\rm Coul},
\label{wb}
\end{equation}
where $E_{\rm vol}=a_{\rm vol}A$ is the volume energy, 
$E_{\rm sym}=a_{\rm sym}[(N-Z)/A]^2 A$ is the symmetry energy,
$E_{\rm surf}=a_{\rm surf}A^{2/3}$ is the surface energy, and 
$E_{\rm Coul}=a_{\rm Coul}Z^2/A^{1/3}$ is the Coulomb energy.
The sum $E_{\rm vol}+E_{\rm sym}$ corresponds to the saturation 
energy of uniform nuclear matter.  Since the matter in a nucleus constitutes
a strongly interacting system, it remains a challenging theoretical problem to 
understand the nuclear matter equation of state (EOS) through microscopic 
calculations that utilize a model of the nuclear force duly incorporating 
low-energy two-nucleon scattering data and properties of very low-mass 
nuclei \cite{HP}.  Furthermore, it is not straightforward to empirically 
clarify the EOS, although constraints on the EOS obtained from nuclear masses 
and radii (e.g., Refs.\ \cite{OTSST,B,CWS,OI}), observables in heavy-ion 
collision experiments performed at intermediate and relativistic energies 
(e.g., Refs.\ \cite{Daniel,LCK}), the isoscalar giant monopole resonance in 
nuclei (e.g., Ref.\ \cite{YCL}), and even X-ray observations of isolated 
neutron stars \cite{Pons} and quiescent low-mass X-ray binaries 
\cite{Guillot} do exist.  In this work we will consider such 
constraints provided by masses of unstable nuclei.

   The energy density of bulk nuclear matter is a function of nucleon density
$n$ and proton fraction $x$, which are related to the neutron and proton number
densities $n_n$ and $n_p$ as $n_n=n(1-x)$ and $n_p=nx$.  We can generally 
express the energy per nucleon near the saturation point of symmetric nuclear
matter as \cite{L}
\begin{equation}
    w=w_0+\frac{K_0}{18n_0^2}(n-n_0)^2+ \left[S_0+\frac{L}{3n_0}(n-n_0)
      \right]\alpha^2.
\label{eos0}
\end{equation}
Here $w_0$, $n_0$ and $K_0$ are the saturation energy, the saturation density 
and the incompressibility of symmetric nuclear matter, and $\alpha=1-2x$ is 
the neutron excess.  $L$ and $S_0$ are associated with the density dependent 
symmetry energy coefficient $S(n)$: $S_0$ is the symmetry energy coefficient 
at $n=n_0$, and $L=3n_0(dS/dn)_{n=n_0}$ is the symmetry energy density 
derivative coefficient (hereafter referred to as the ``density symmetry 
coefficient'').  As the neutron excess increases from zero, the saturation 
point moves in the density versus energy plane.  This movement is determined 
mainly by the parameters $L$ and $S_0$.  Up to second order in $\alpha$, the 
saturation energy $w_s$ and density $n_s$ are given by 
\begin{equation}
  w_s=w_0+S_0 \alpha^2
\label{ws}
\end{equation}
and
\begin{equation}
  n_s=n_0-\frac{3 n_0 L}{K_0}\alpha^2.
\label{ns}
\end{equation}
The slope, $y$, of the saturation line near $\alpha=0$ $(x=1/2)$ is thus 
expressed as
\begin{equation}
 y=-\frac{K_0 S_0}{3 n_0 L}.
\label{slope}
\end{equation}

      In our earlier investigations \cite{OI} we explored a systematic way of 
extracting $L$ and $S_0$ from empirical masses and radii of nuclei, together 
with the parameters, $n_0$, $w_0$ and $K_0$, characterizing the saturation of 
symmetric nuclear matter.  We first gave an expression for the energy of 
uniform nuclear matter, which reduces to the phenomenological form 
(\ref{eos0}) in the simultaneous limit $n\to n_0$ and $\alpha\to0$ 
$(x\to 1/2)$.  Using this energy expression within a simplified version of 
the extended Thomas-Fermi approximation, which permits us to determine the 
macroscopic features of the nuclear ground state, we calculated charges, 
charge radii and masses of $\beta$-stable nuclei for fixed $A$.  Comparing 
these calculations with empirical values allows us to derive the optimal 
parameter set for fixed values of the slope $y$ and the incompressibility 
$K_0$.  We thus found a strong correlation between $L$ and $S_0$.  The next 
step was to calculate the root-mean-square (rms) charge and matter radii of 
more neutron-rich nuclei that are expected to be produced in 
radioactive ion beam facilities.  The results suggest that the density 
symmetry coefficient $L$ may be constrained by possible systematic data for 
the matter radii in a manner that is nearly independent of $K_0$.

     The reason that we concentrated on radii of unstable nuclei rather than
their masses in Ref.\ \cite{OI} was that we originally considered the 
surface and electrostatic properties as well as shell and pairing effects, 
which play a role in nuclear masses, to obscure the derivation of the EOS 
parameters.  Nevertheless, in the present analysis, we perform a systematic 
calculation of nuclear masses from the same framework used for our previous 
calculations of nuclear radii.  Since experimental mass data have been 
accumulated even for unstable nuclei, we can compare the calculations with 
the existing data, which is a great advantage over the case of nuclear radii.
Comparison of the two-proton separation energy 
($S_{2p}(Z,N)=E_B(Z,N)-E_B(Z-2,N)$) implies that a larger $L$, corresponding to
smaller symmetry energy at subnuclear densities, is favored.  This 
tendency is most clearly elucidated from the masses of light, nondeformed, 
and neutron-rich nuclei.  This can be understood from a compressible 
liquid-drop picture of nuclei in terms of the $L$ dependence of the surface 
symmetry energy, which has some relevance to the neutron skin thickness 
\cite{IO,Warda}.

     In Sec.\ II we summarize a macroscopic model of nuclei used here.
Calculations of nuclear masses and the comparison with empirical data
are illustrated in Sec.\ III.  Our conclusions are presented in Sec.\ IV.

\section{Macroscopic nuclear model}
\label{sec:model}

     In this section, we summarize a macroscopic model of nuclei 
\cite{OI}, which was constructed in such a way as to reproduce the known 
global properties of stable nuclei and can be used for describing the masses 
and radii of unstable nuclei in a manner that is dependent on the EOS of 
nuclear matter.

    The bulk energy per nucleon is an essential ingredient of the
macroscopic nuclear model.  We set this energy as 
\begin{eqnarray}
  w&=&\frac{3 \hbar^2 (3\pi^2)^{2/3}}{10m_n n}(n_n^{5/3}+n_p^{5/3})
   \nonumber \\ & & +(1-\alpha^2)v_s(n)/n+\alpha^2 v_n(n)/n,
\label{eos1}
\end{eqnarray}
where 
\begin{equation}
  v_s=a_1 n^2 +\frac{a_2 n^3}{1+a_3 n}
\label{vs}
\end{equation}
and
\begin{equation}
  v_n=b_1 n^2 +\frac{b_2 n^3}{1+b_3 n}
\label{vn}
\end{equation}
are the potential energy densities for symmetric nuclear matter and pure 
neutron matter, $n_n$ and $n_p$ are the neutron and proton number densities,
$n=n_n+n_p$, $\alpha=(n_n-n_p)/n$ is the neutron excess, and $m_n$ is the 
neutron mass.  Expressions (\ref{eos1})--(\ref{vn}) can well reproduce the 
microscopic calculations of symmetric nuclear matter and pure neutron matter 
by Friedman and Pandharipande \cite{FP} in the variational method.  In this 
method, the isospin dependence of asymmetric matter EOS is shown to be well 
approximated by Eq.\ (\ref{eos1}) \cite{LP}.  (Replacement of the proton mass 
$m_p$ by $m_n$ in the proton kinetic energy would make only a negligible 
difference.)  For the later purpose of roughly describing the nucleon 
distribution in a nucleus, we incorporate into the potential energy densities 
(\ref{vs}) and (\ref{vn}) a low density behavior $\propto n^2$ as expected 
from a contact two-nucleon interaction.  
A set of expressions (\ref{eos1})--(\ref{vn}) is one of the simplest 
that reduces to the usual form (\ref{eos0}) in the limit of $n\to n_0$ and 
$\alpha\to0$.  In fact, the parameters $a_1, \cdots, b_3$ are related to
$n_0$, $w_0$, $K_0$, $S_0$, and $L$ as
\begin{equation}
 S_0= \frac16 \left(\frac{3\pi^2}{2}\right)^{2/3}\frac{\hbar^2}{m_n}n_0^{2/3}
  +(b_1-a_1)n_0+\left(\frac{b_2}{1+b_3 n_0}-\frac{a_2}{1+a_3 n_0}\right)n_0^2,
\label{s0}
\end{equation}
\begin{eqnarray}
 \frac13 n_0 L
 &=&\frac19\left(\frac{3\pi^2}{2}\right)^{2/3}\frac{\hbar^2}{m_n}n_0^{5/3}
     +(b_1-a_1)n_0^2
 +2\left(\frac{b_2}{1+b_3 n_0}-\frac{a_2}{1+a_3 n_0}\right)n_0^3
\nonumber \\ & &
     -\left[\frac{b_2 b_3}{(1+b_3 n_0)^2}
           -\frac{a_2 a_3}{(1+a_3 n_0)^2}\right]n_0^4,
\label{p0}
\end{eqnarray}
\begin{equation}
 w_0=\frac{3}{10}\left(\frac{3\pi^2}{2}\right)^{2/3}\frac{\hbar^2}{m_n}
     n_0^{2/3} + a_1 n_0+\frac{a_2 n_0^2}{1+a_3 n_0},
\label{w0}
\end{equation}
\begin{equation}
 K_0=-\frac35\left(\frac{3\pi^2}{2}\right)^{2/3}\frac{\hbar^2}{m_n}n_0^{2/3}
     +\frac{18 a_2 n_0^2}{(1+a_3 n_0)^3},
\label{k0}
\end{equation}
\begin{equation}
 0=\frac15\left(\frac{3\pi^2}{2}\right)^{2/3}\frac{\hbar^2}{m_n}n_0^{-1/3}
     +a_1+\frac{2a_2 n_0}{1+a_3 n_0}-\frac{a_2 a_3 n_0^2}{(1+a_3 n_0)^2}.
\label{sat}
\end{equation}

    We determine the parameters $a_1, \cdots, b_3$ in such a way that
the charge number, charge radius, and mass of stable nuclei calculated in a 
macroscopic nuclear model constructed in Ref.\ \cite{OI} are consistent with 
the empirical data.  In the course of this determination, we fix $b_3$, 
which controls the EOS of matter for large neutron excess and high density,
at 1.58632 fm$^3$.  This value was obtained by one of the authors \cite{O} in 
such a way as to reproduce the neutron matter energy of Friedman and 
Pandharipande \cite{FP}.  Change in this parameter would make no significant
difference in the determination of the other parameters and the final 
results for nuclear masses.

    We describe macroscopic nuclear properties in a way dependent on the EOS
parameters $a_1, \cdots, b_3$ by using a Thomas-Fermi model \cite{OI}.  The 
essential point of this model is to write down the total energy of a nucleus 
of mass number $A$ and charge number $Z$
as a function of the density distributions $n_n({\bf r})$ and $n_p({\bf r})$ 
in the form
\begin{equation}
 E=E_b+E_g+E_C+Nm_n c^2+Zm_p c^2,
\label{e}
\end{equation}
where 
\begin{equation}
  E_b=\int d^3 r n({\bf r})w\left(n_n({\bf r}),n_p({\bf r})\right)
\label{eb}
\end{equation}
is the bulk energy,
\begin{equation}
  E_g=F_0 \int d^3 r |\nabla n({\bf r})|^2
\label{eg}
\end{equation}
is the gradient energy with adjustable constant $F_0$,
\begin{equation}
  E_C=\frac{e^2}{2}\int d^3 r \int  d^3 r' 
      \frac{n_p({\bf r})n_p({\bf r'})}{|{\bf r}-{\bf r'}|}
\label{ec}
\end{equation}
is the Coulomb energy, and $N=A-Z$ is the neutron number.  This functional 
allows us to connect the EOS and the nuclear binding energy through the bulk 
energy part $E_b$.  For simplicity we use the following parametrization for 
the nucleon distributions $n_i(r)$ $(i=n,p)$: 
\begin{equation}
  n_i(r)=\left\{ \begin{array}{lll}
  n_i^{\rm in}\left[1-\left(\displaystyle{\frac{r}{R_i}}\right)^{t_i}\right]^3,
         & \mbox{$r<R_i,$} \\
             \\
         0,
         & \mbox{$r\geq R_i.$}
 \end{array} \right.
\label{ni}
\end{equation}
This parametrization allows for the central density, half-density radius, and 
surface diffuseness for neutrons and protons separately.

     In order to construct the nuclear model in such a way as to reproduce 
empirical masses and
radii of stable nuclei, we first extremized the binding energy with respect to
the particle distributions for fixed mass number, five EOS parameters 
$a_1, \cdots, b_2$, and gradient coefficient $F_0$.  
Next, for various sets of the incompressibility $K_0$ and 
the density symmetry coefficient $L$, we obtained the remaining three EOS 
parameters $n_0$, $w_0$, and $S_0$ and the gradient coefficient $F_0$
by fitting the calculated optimal 
values of charge number, mass excess, rms charge radius to 
empirical data for stable nuclei on the smoothed $\beta$ stability line 
\cite{O} and by using Eqs.\ (\ref{s0})--(\ref{sat}).
In the range of the parameters 
$0<L<160$ MeV and 180 MeV $<K_0<360$ MeV, as long as $y \lesssim -200$ MeV 
fm$^3$, we obtained a reasonable fitting to such data (see Fig.\ 1).
As a result of this fitting, the parameters $n_0$, $w_0$, $S_0$, and $F_0$
are constrained as $n_0=0.14$--0.17 fm$^{-3}$, $w_0=-16\pm1$ MeV, 
$S_0=25$--40 MeV, and $F_0=66\pm6$ MeV fm$^5$.
We remark that a negative $L$ is inconsistent with the fact that the size
of $A=17,20,31$ isobars deduced from the experimental values of the 
interaction cross section tends to increase with
neutron/proton excess \cite{isobar}.  This inconsistency can be seen from 
Eq.\ (\ref{ns}) which shows that the saturation density $n_s$ 
increases (and hence the isobar size decreases) 
with neutron/proton excess for a negative $L$.  
We also note that the fitting gives rise to a relation nearly independent
of $K_0$,
\begin{equation}
  S_0\approx B+CL,
  \label{linear}
\end{equation}
where $B\approx28$ MeV and $C\approx0.075$.  As we shall see, this linear 
relation plays a part in the $L$ dependence of calculated masses at large
neutron excess.

\begin{figure}[t]
\includegraphics[width=8.5cm]{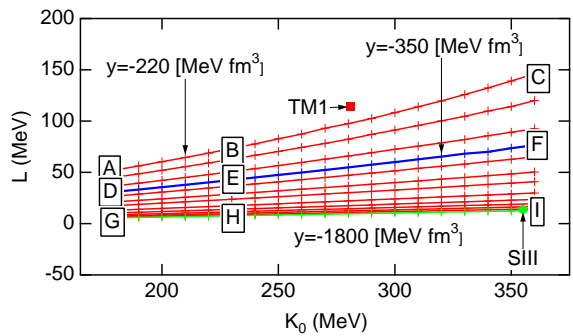}
\vspace{-0.5cm}
\caption{\label{lk0} (Color online)
The sets of $(L,K_0)$ (crosses) consistent with the mass 
and radius data for stable nuclei.  The thin lines are lines of constant $y$.
The labels A--I denote the sets for which we performed detailed 
calculations of the ground state properties of inhomogeneous nuclear matter 
at subnuclear densities \cite{OI2}. In this paper we often focus on the
EOS models with the parameter sets C and G.
   For comparison, the values calculated from two mean-field models 
[TM1 (square) and SIII (dot)], which are known to be extreme cases 
\cite{OTSST}, are plotted.  The plot shows that our sets of $(L,K_0)$
effectively cover such extreme cases and constraints on $L$ and $K_0$
from other observables \cite{constraints}.}
\end{figure}

     We remark that in the range of the EOS parameters $L$ and $K_0$
shown in Fig.\ 1, the calculations agree well with a more 
extended data set of nuclear masses for $A\geq2$ \cite{Audi} and charge radii 
for $A\geq50$ \cite{dV}.  The rms deviations of the calculated masses from the 
measured values are $\sim3$--5 MeV, which are comparable with the deviations 
obtained from a Weizs{\" a}cker-Bethe formula, while the rms deviations of the 
calculated charge radii from the measured values are about 0.06 fm, which 
are comparable with the deviations obtained from the $A^{1/3}$ law.
As we shall see, detailed comparison with empirical masses in terms of
two-proton separation energy allows us to tell which of the EOS models with 
large $L$ and with small $L$ are more favored.

\section{Nuclear masses}
\label{sec:mass}

     We now proceed to evaluate nuclear masses from various EOS models
with the parameter set $(L,K_0)$ consistent with the mass and radius data 
for stable nuclei by minimizing Eq.\ (\ref{e}) for fixed $N$ and $Z$.
The results are then compared with the existing data
in terms of the two-proton separation energy.  We finally discuss the 
$L$ dependence of the calculated masses within the framework of a compressible 
liquid-drop picture of nuclei.


     We begin by illustrating the calculated and experimental values of the
two-proton separation energy $S_{2p}$ for O, Mg, Ca, Ni, Sn, and Pb isotopes.  
The two-proton separation energy is useful partly because the even-odd
staggering is essentially cancelled out and partly because the isotope 
dependence except for shell gaps is smooth according to the  
Yamada-Matumoto systematics \cite{YM}.  In fact, as can be seen from Fig.\ 2 
in which the two-proton separation energy minus the one calculated from 
a Weizs{\" a}cker-Bethe mass formula (\ref{wb}) with the coefficients 
$a_{\rm vol}=-15.5391$ MeV, $a_{\rm sym}=22.7739$ MeV, $a_{\rm surf}=16.9666$ 
MeV, and $a_{\rm Coul}=0.703893$ MeV \cite{Y},
i.e., $S_{2p}-S_{2p}^{\rm WB}$, is plotted, the empirical 
values behave monotonously with neutron excess except for the vicinity of 
$N=Z$ and neutron magic numbers and for the deformed region.  Note that 
this monotonous behavior is clearer for O, Mg, and Ca than for Ni, Sn, 
and Pb.

\begin{figure}[t]
 \begin{minipage}{.45\linewidth}
  \includegraphics[width=8.5cm]{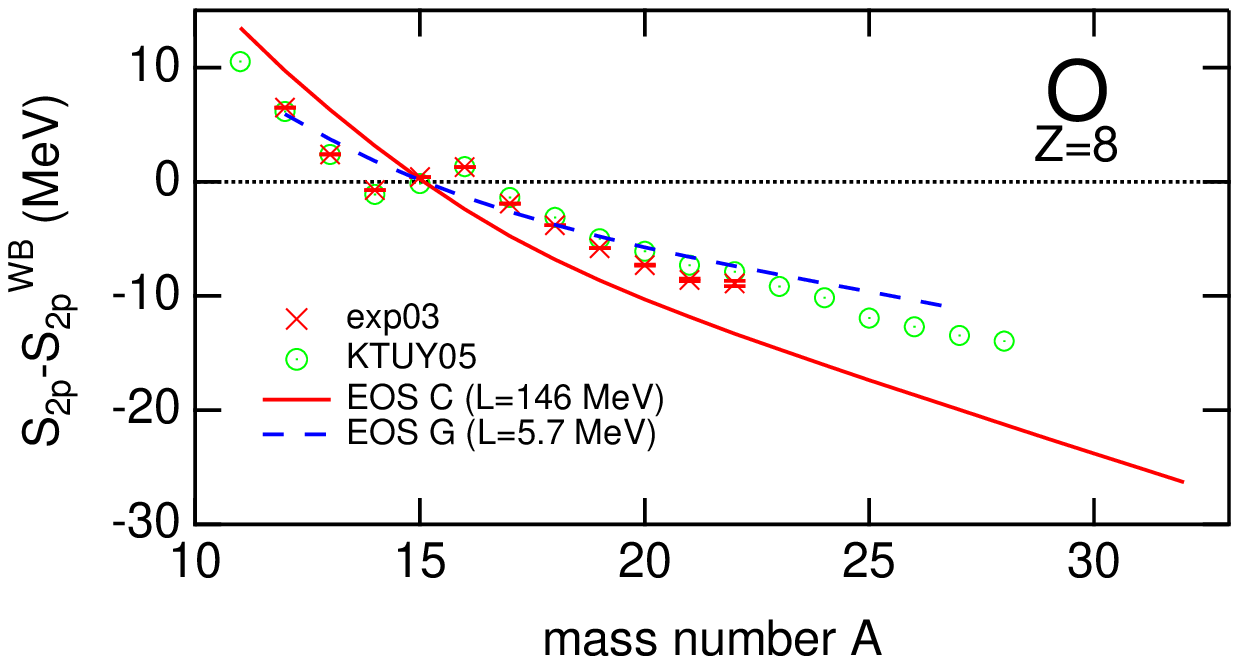}  
 \end{minipage}
 \begin{minipage}{.45\linewidth}
  \includegraphics[width=8.5cm]{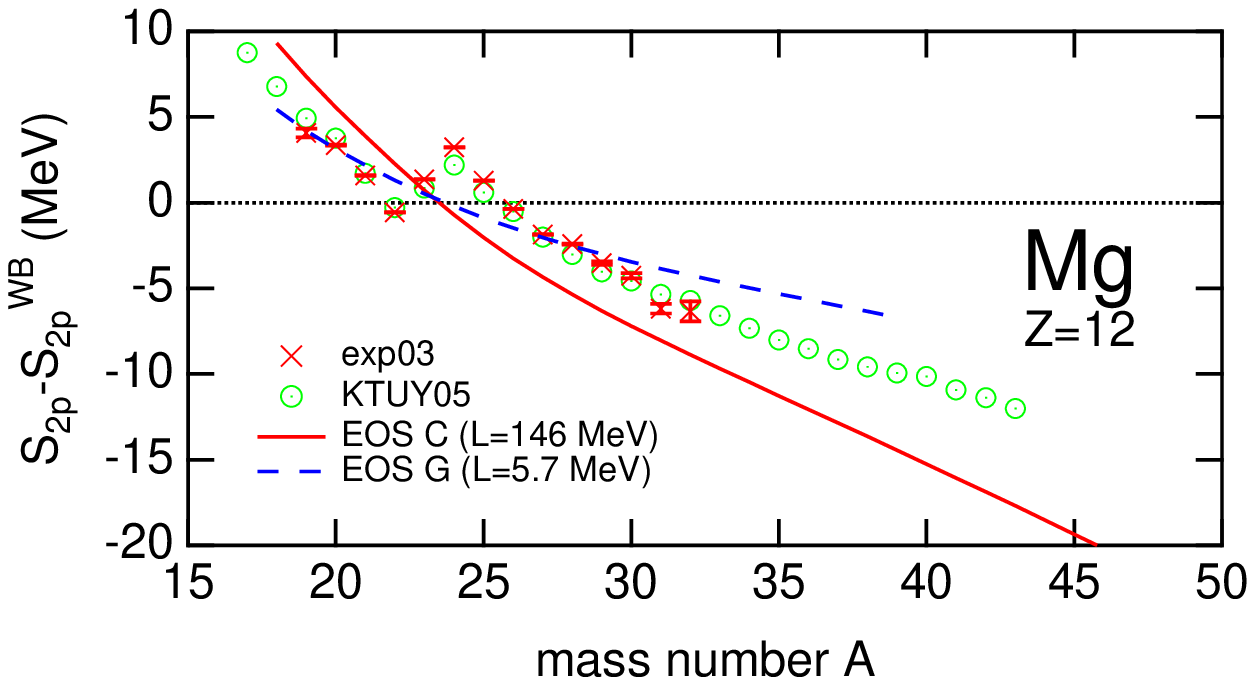}
 \end{minipage}
 \begin{minipage}{.45\linewidth}
  \includegraphics[width=8.5cm]{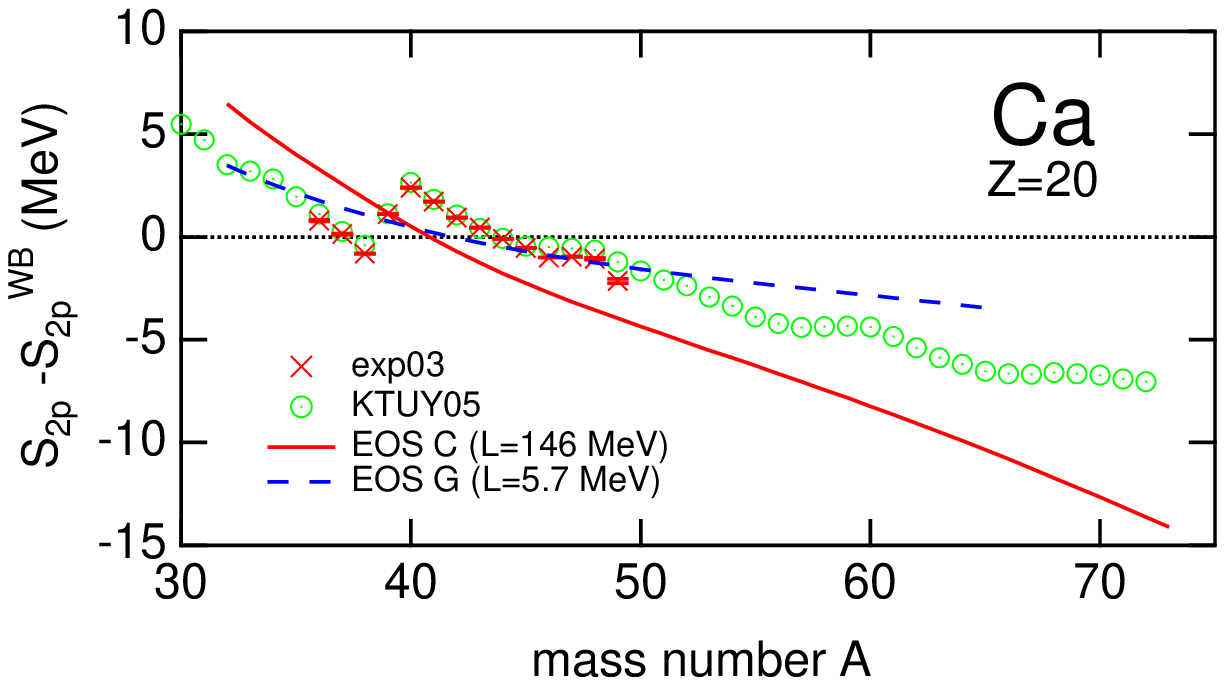}
 \end{minipage}
 \begin{minipage}{.45\linewidth}
  \includegraphics[width=8.5cm]{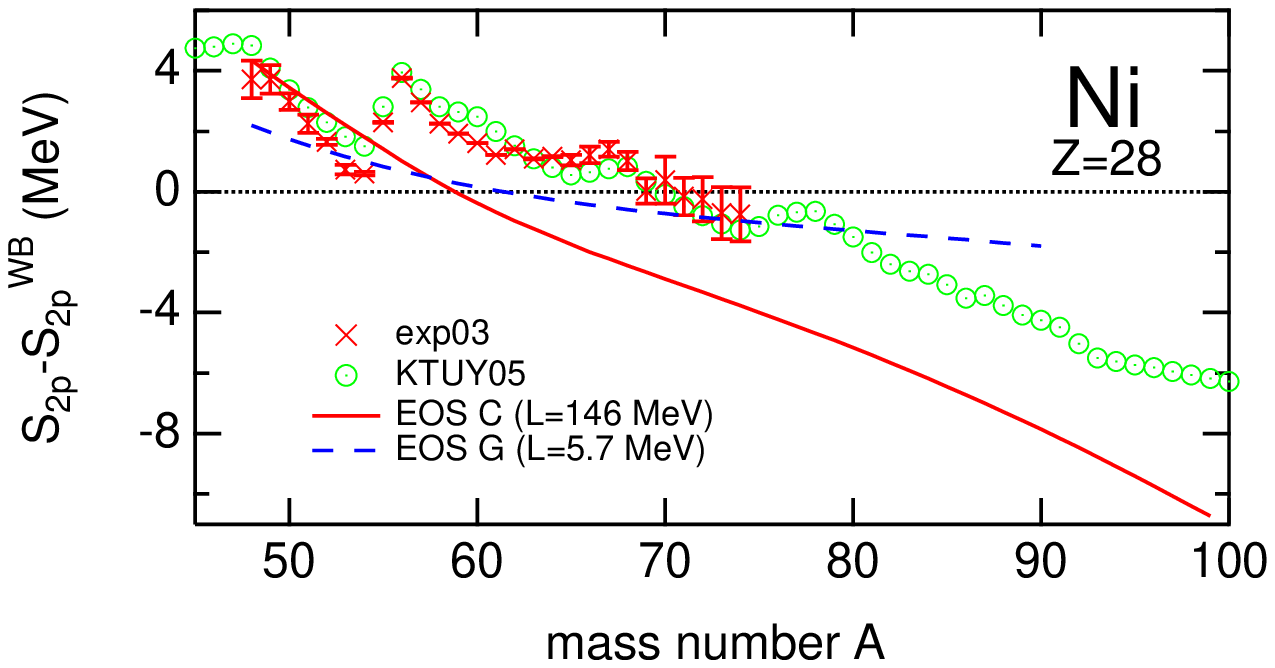}
 \end{minipage}
 \begin{minipage}{.45\linewidth}
  \includegraphics[width=8.5cm]{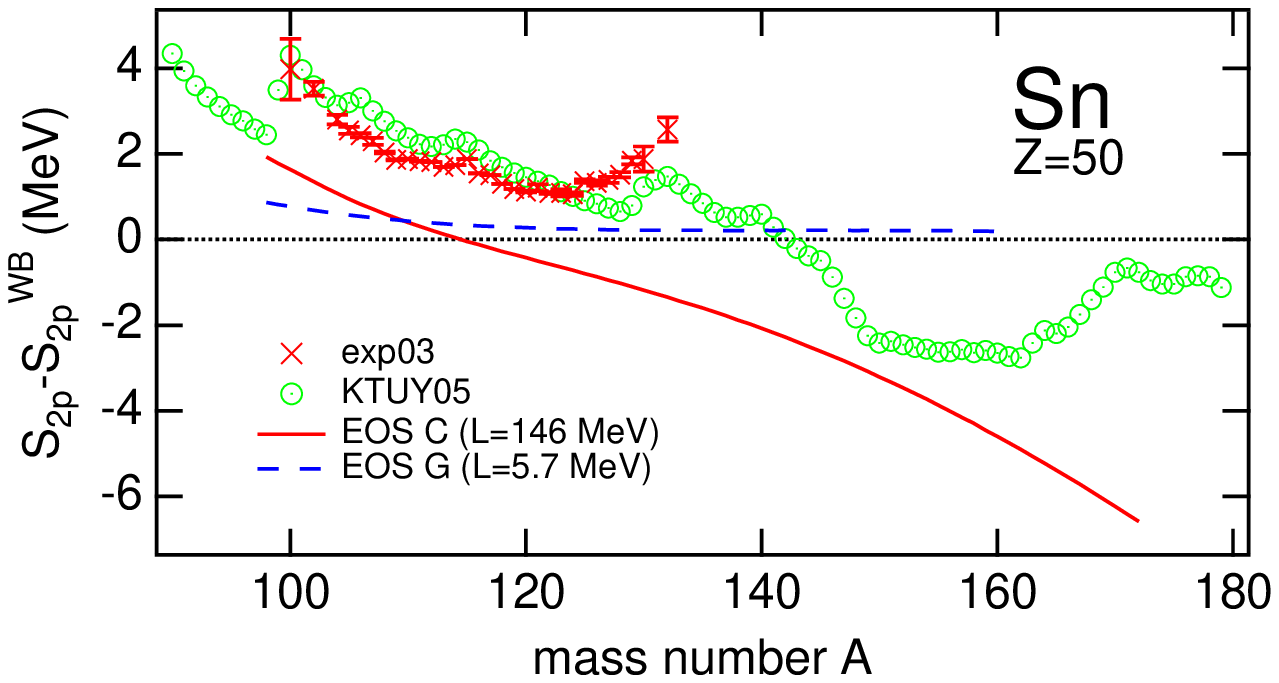}
 \end{minipage}
 \begin{minipage}{.45\linewidth}
  \includegraphics[width=8.5cm]{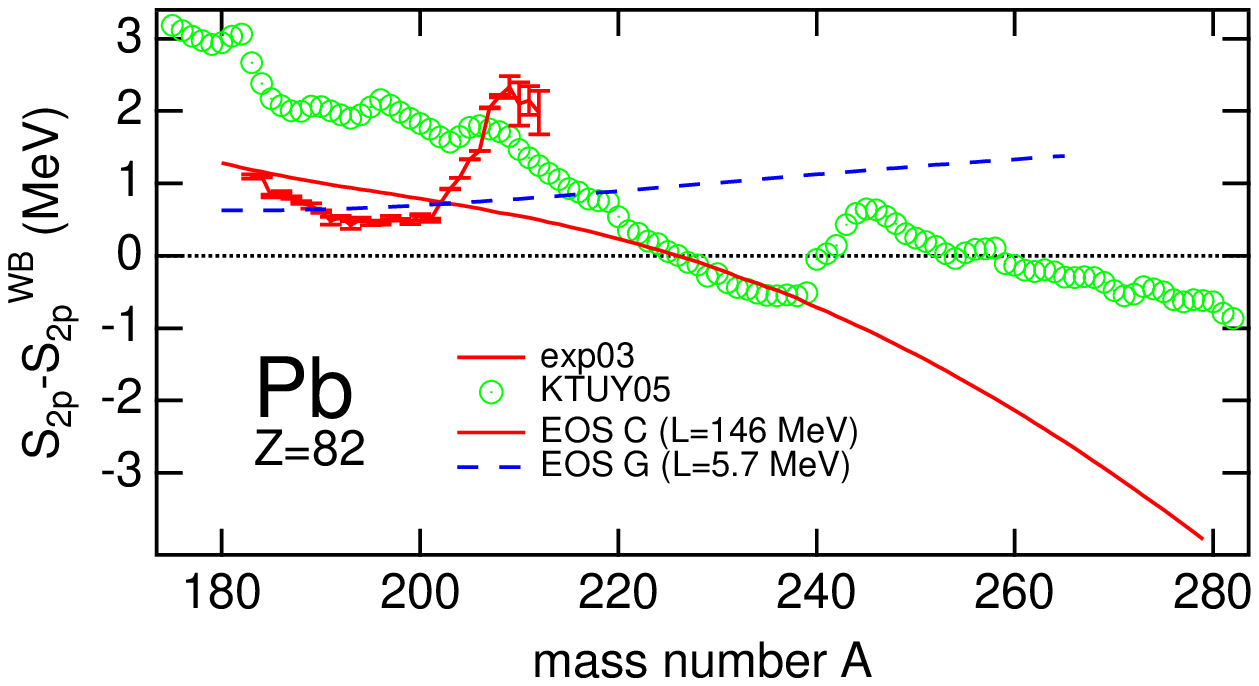}
 \end{minipage}
 \caption{\label{S2p} (Color online)
The two-proton separation energy, having the one calculated from a 
Weizs{\" a}cker-Bethe mass formula \cite{Y} subtracted out, for O, Mg, Ca, Ni, 
Sn, and Pb isotopes.  The empirical values \cite{Audi}, the calculated values 
from the EOS models G and C, and the values obtained from a contemporary mass 
formula \cite{KTUY} are plotted in each panel.}
\end{figure}

     We remark that the macroscopic nuclear model summarized in the 
previous section was originally used for systematic calculations of 
charge and matter radii of nuclei of $A\gtrsim50$ \cite{OI}.  Here, we 
performed systematic calculations of masses of nuclei including O, Mg, and
Ca isotopes.  Applicability to lighter nuclei can be confirmed from Fig.\ 2,
which shows that differences between the calculated and empirical values of 
$S_{2p}$ for stable nuclei are limited within 3 MeV.

     The calculated values of $S_{2p}$, neither including 
deformation, the Wigner term, nor shell corrections, show a smooth dependence 
on neutron excess in a way different between the EOS models G ($L=5.7$ MeV) 
and C ($L=146$ MeV). (The Weizs{\" a}cker-Bethe formula, which is based on the 
incompressible liquid-drop model for nuclei, is independent of the parameters 
$L$ and $K_0$ characterizing the density dependence of the EOS.)  In order to 
examine how the calculated values of $S_{2p}$ depend on $L$ and $K_0$, we plot 
in Fig.\ 3 the results obtained for $^{78}$Ni and $^{22}$O
from the EOS models with various 
values of $L$ and $K_0$ shown in Fig.\ 1.  The results decrease with $L$ 
almost linearly, while they are nearly independent of $K_0$.  Accordingly, 
for various values of $L$ one can predict where $S_{2p}-S_{2p}^{\rm WB}$ is 
located in Fig.\ 2 by interpolation or extrapolation of the results from the 
EOS models G and C.  As far as the slope of $S_{2p}-S_{2p}^{\rm WB}$ is 
concerned, the result with the EOS model C seems to be more consistent with 
the empirical behavior, particularly for light nuclei.

\begin{figure}[t]
 \begin{minipage}{.45\linewidth}
  \includegraphics[width=8.5cm]{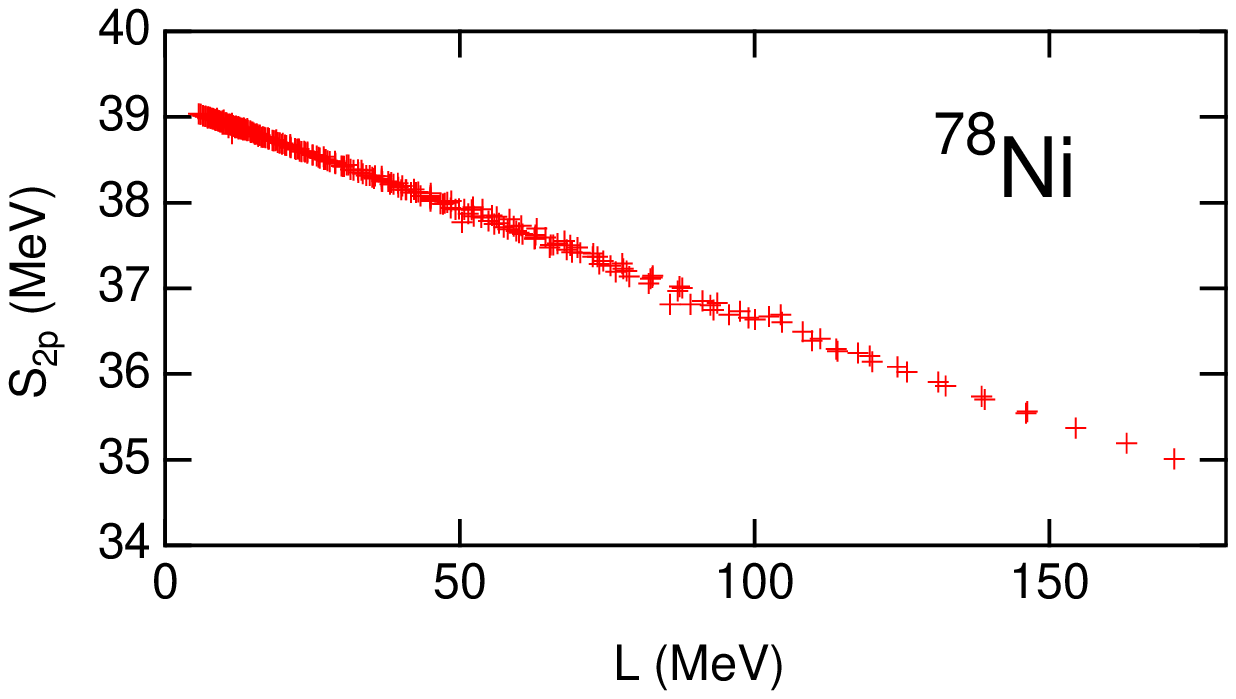}
 \end{minipage}
 \begin{minipage}{.45\linewidth}
  \includegraphics[width=8.5cm]{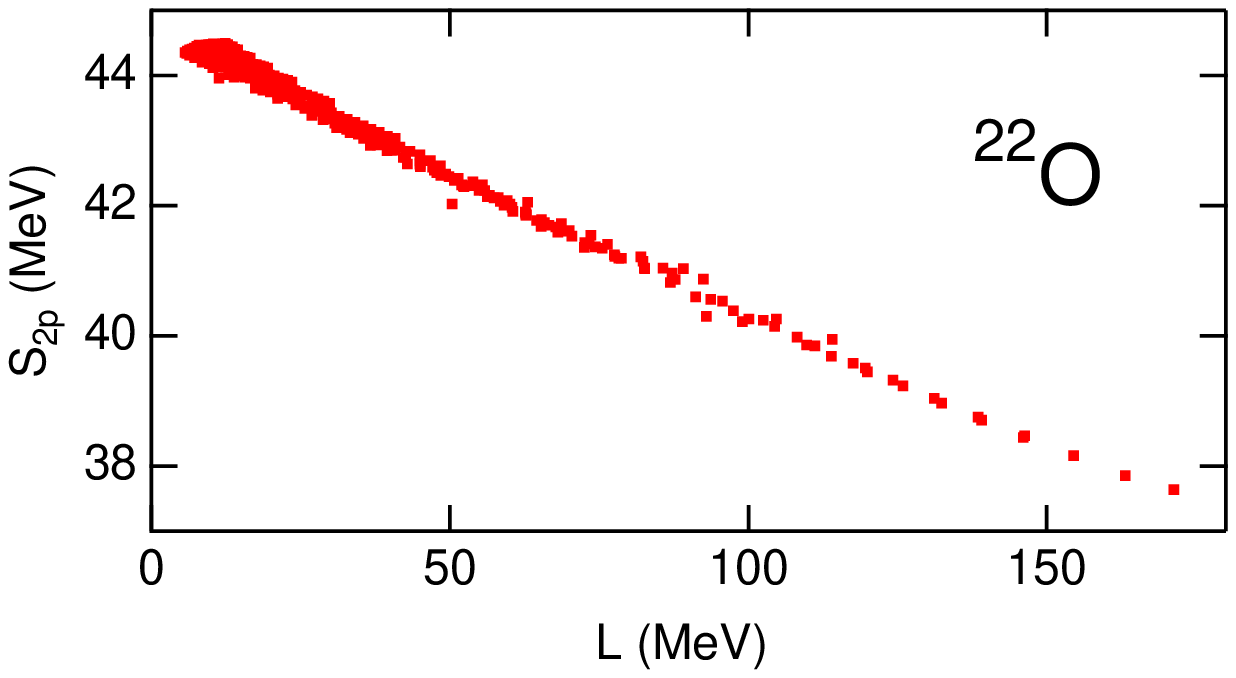}
 \end{minipage}
\vspace{-0.5cm}
\caption{\label{S2pNi} (Color online)
The two-proton separation energy calculated for $^{78}$Ni 
and $^{22}$O as a function of $L$.
}
\end{figure}

     We also notice from Fig.\ 2 that there is a roughly uniform offset 
between the empirical values and the values calculated from the EOS model C
at $N>Z$.  This offset comes mainly from the proton shell gaps.  In fact,
it is effectively cancelled out in a region away from $Z=N$ and neutron
magic numbers by taking a difference, 
$S_{2p}(Z,N)-S_{2p}(Z,N-2)\equiv4\delta V_{np}$, as can be seen from 
Fig.\ 4.  In contrast, the empirical behavior of this difference 
shows a $N=Z$ and neutron shell structure in a more exaggerated manner.
As discussed in Ref.\ \cite{Stoitsov}, the smoothed behavior of 
$\delta V_{np}$ is related to the coefficients $a_{\rm sym}$ and 
$a_{\rm ssym}$ affixed to 
$A[(N-Z)/A]^2$ and $A^{2/3}[(N-Z)/A]^2$ in the mass formula (\ref{wb}) 
(with the surface symmetry term added) by
\begin{equation}
\delta V_{np}\approx 2(a_{\rm sym}+a_{\rm ssym}A^{-1/3})/A.
\label{vpn}
\end{equation}

\begin{figure}[t]
 \begin{minipage}{.45\linewidth}
  \includegraphics[width=8.5cm]{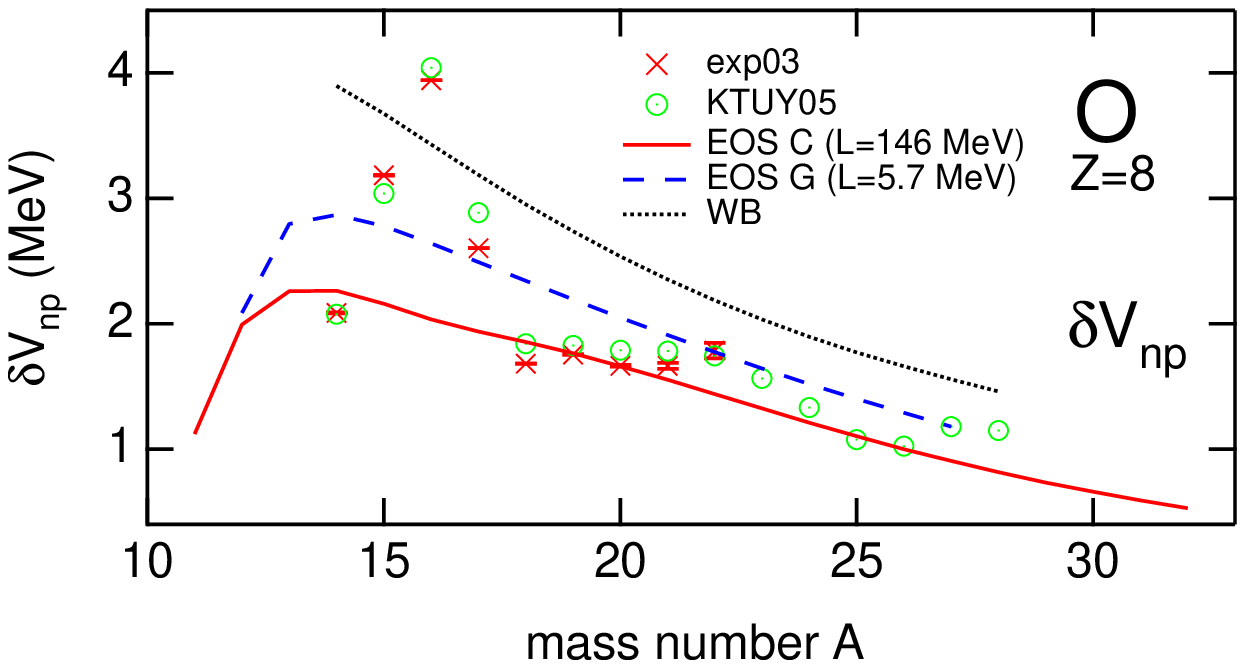}  
 \end{minipage}
 \begin{minipage}{.45\linewidth}
  \includegraphics[width=8.5cm]{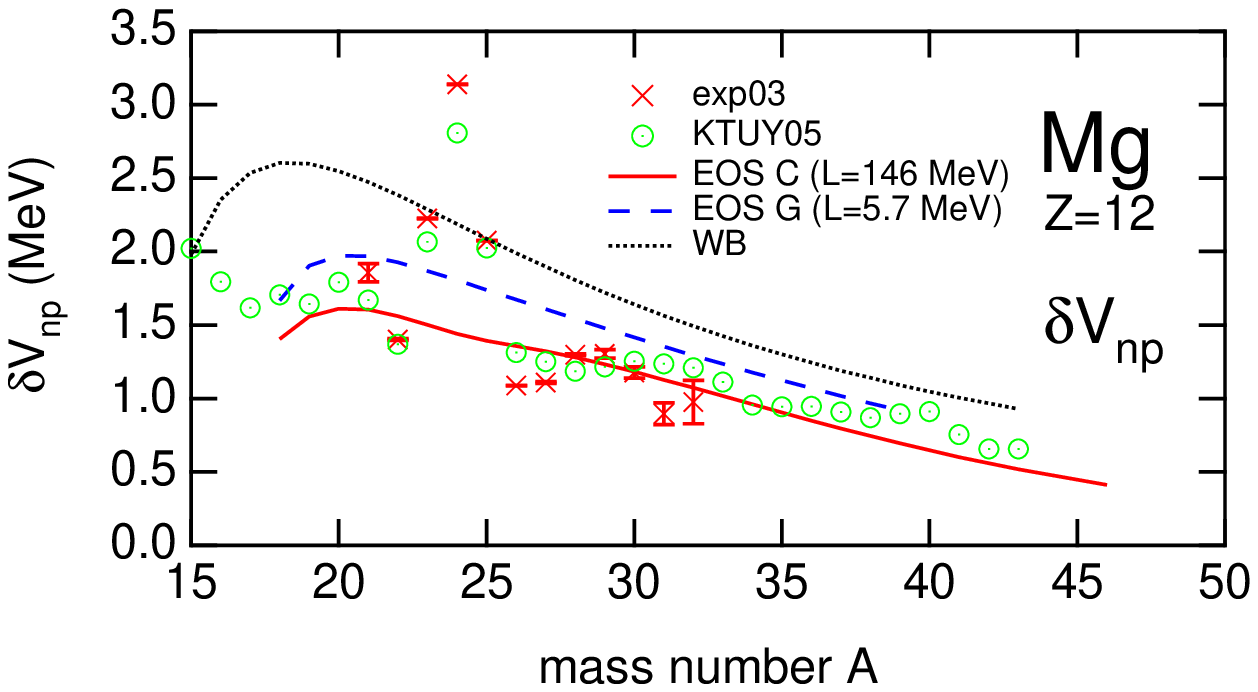}
 \end{minipage}
 \begin{minipage}{.45\linewidth}
  \includegraphics[width=8.5cm]{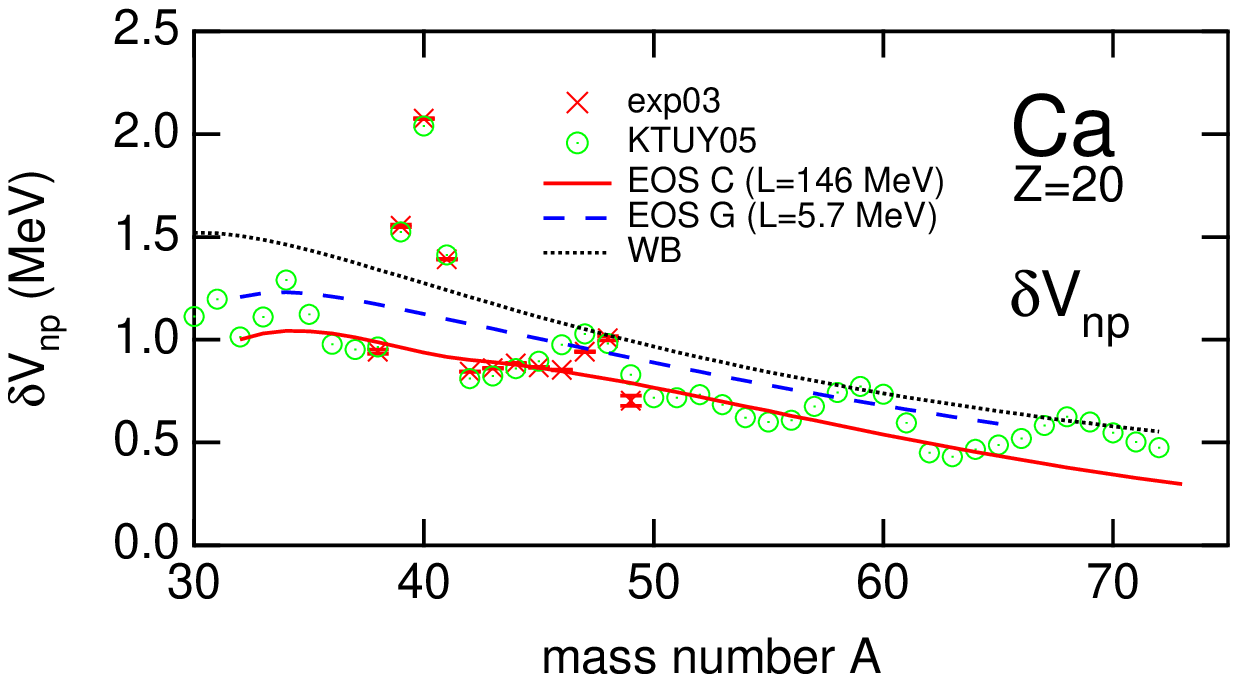}
 \end{minipage}
 \begin{minipage}{.45\linewidth}
  \includegraphics[width=8.5cm]{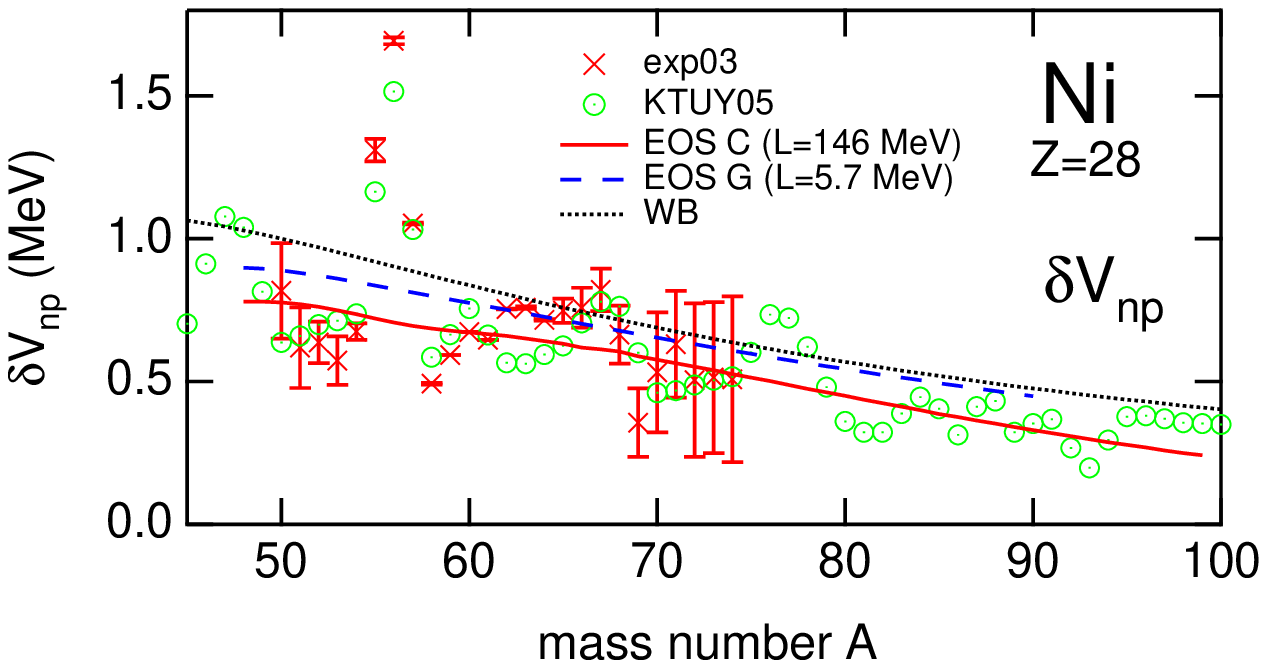}
 \end{minipage}
 \begin{minipage}{.45\linewidth}
  \includegraphics[width=8.5cm]{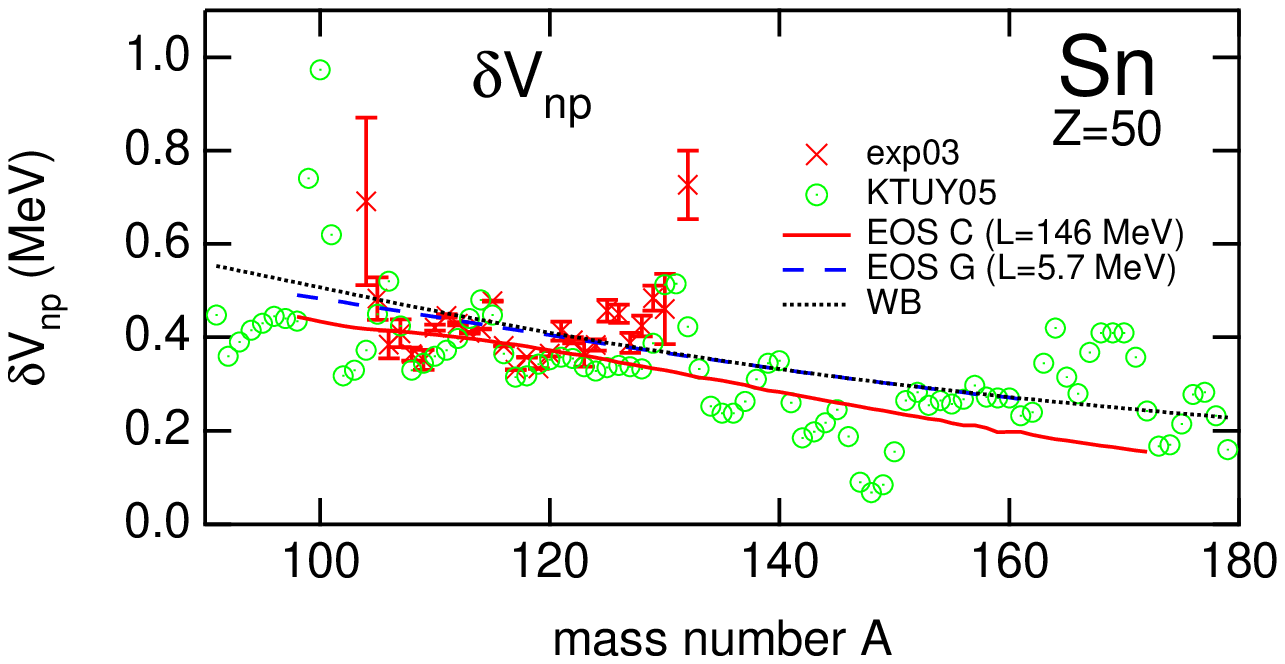}
 \end{minipage}
 \begin{minipage}{.45\linewidth}
  \includegraphics[width=8.5cm]{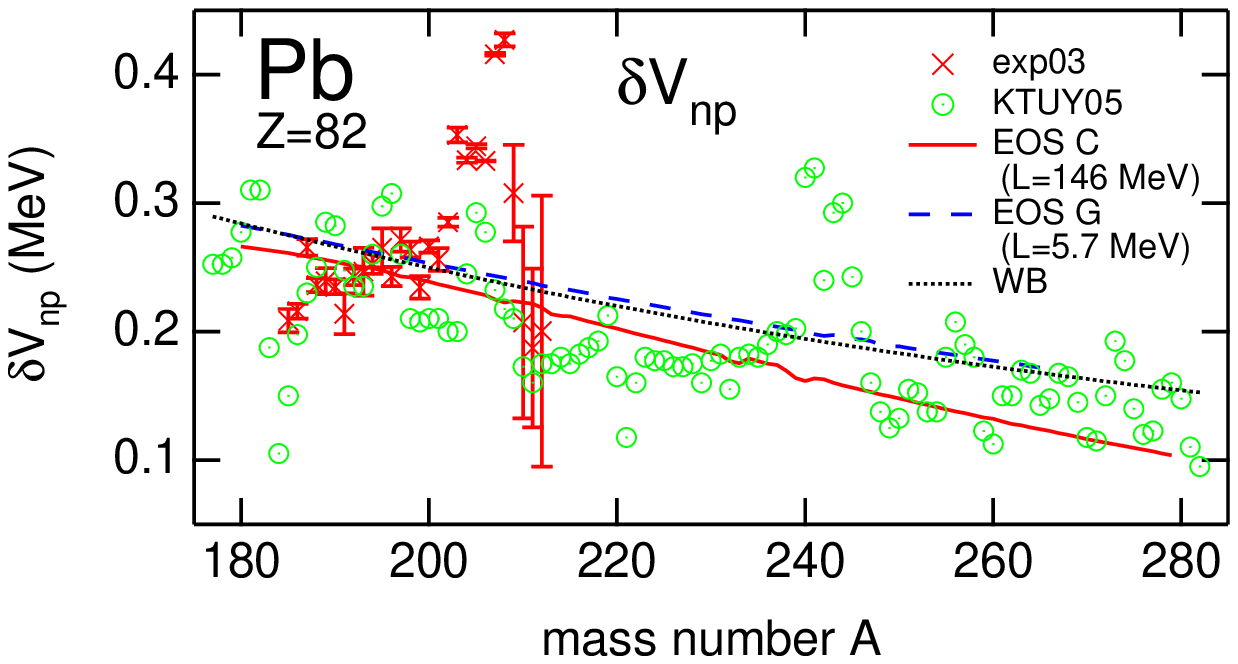}
 \end{minipage}
\caption{\label{vnp} (Color online)
$\delta V_{np}$ for O, Mg, Ca, Ni, Sn, and Pb isotopes.  In each panel,
the empirical values \cite{Audi}, the values from a contemporary
mass formula \cite{KTUY} and a Weizs{\" a}cker-Bethe mass formula 
\cite{Y}, and the calculated values from the EOS
models C and G are plotted. 
}
\end{figure}

      Our results for $\delta V_{np}$ suggest that the $L$ dependence
lies in the parameters $a_{\rm sym}$ and $a_{\rm ssym}$.  In fact, this 
suggestion can be understood within the framework of a compressible
liquid-drop model in which nuclei in equilibrium are allowed to 
have a density different from the saturation density $n_0$ of
symmetric nuclear matter.  If one ignores Coulomb and surface corrections,
the equilibrium density and energy per nucleon of a liquid-drop 
correspond to $n_s$ and $w_s$, which are given by Eqs.\ (\ref{ns}) 
and (\ref{ws}) for nearly symmetric nuclear matter.  Note that
the parameter $a_{\rm sym}$ is characterized by the symmetry energy
coefficient $S_0$, which in turn is related to $L$ by Eq.\ (\ref{linear}).
Since one obtains a larger $a_{\rm sym}$ for larger $L$, the effect of
$a_{\rm sym}$ tends to increase $\delta V_{np}$ with $L$, which is
in the wrong direction.  This effect, therefore, has to be dominated 
by the effect of $a_{\rm ssym}$.  This is consistent with the fact that
the $L$ dependence of the calculated $S_{2p}$ is clearer for lighter 
nuclei and also with the tendency that the calculated mass decreases
with $L$ for neutron-rich nuclei (see Fig.\ 5).

\begin{figure}[t]
 \begin{minipage}{.45\linewidth}
  \includegraphics[width=8.5cm]{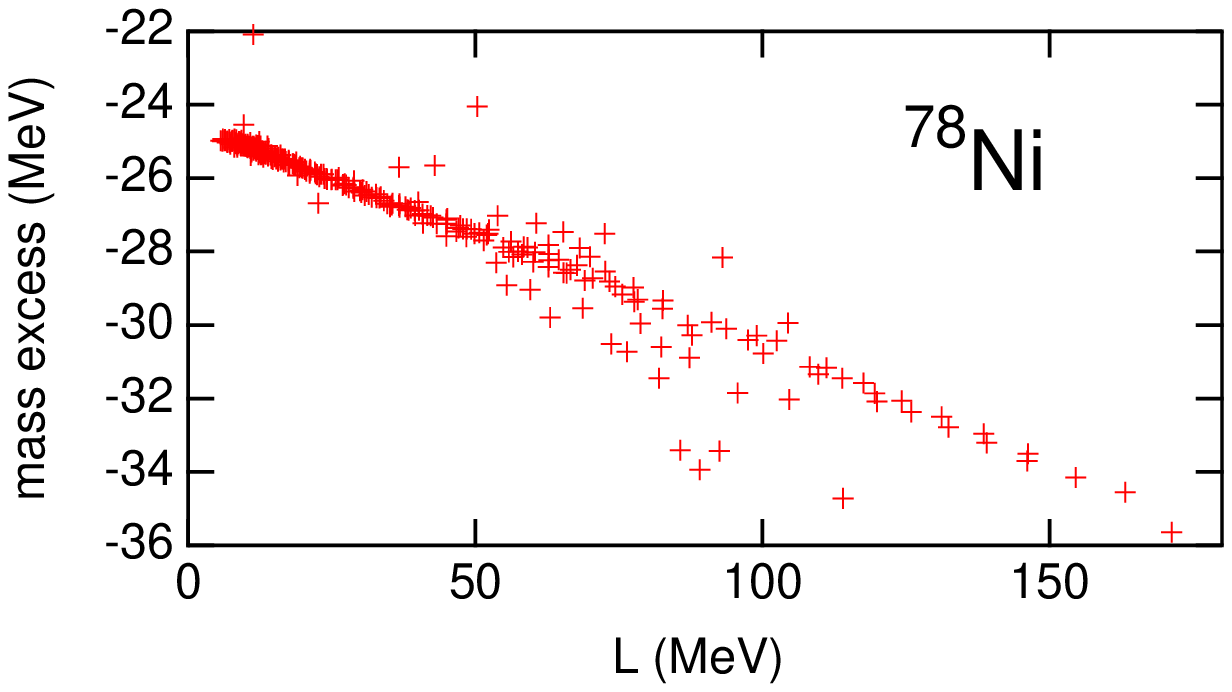}
 \end{minipage}
 \begin{minipage}{.45\linewidth}
  \includegraphics[width=8.5cm]{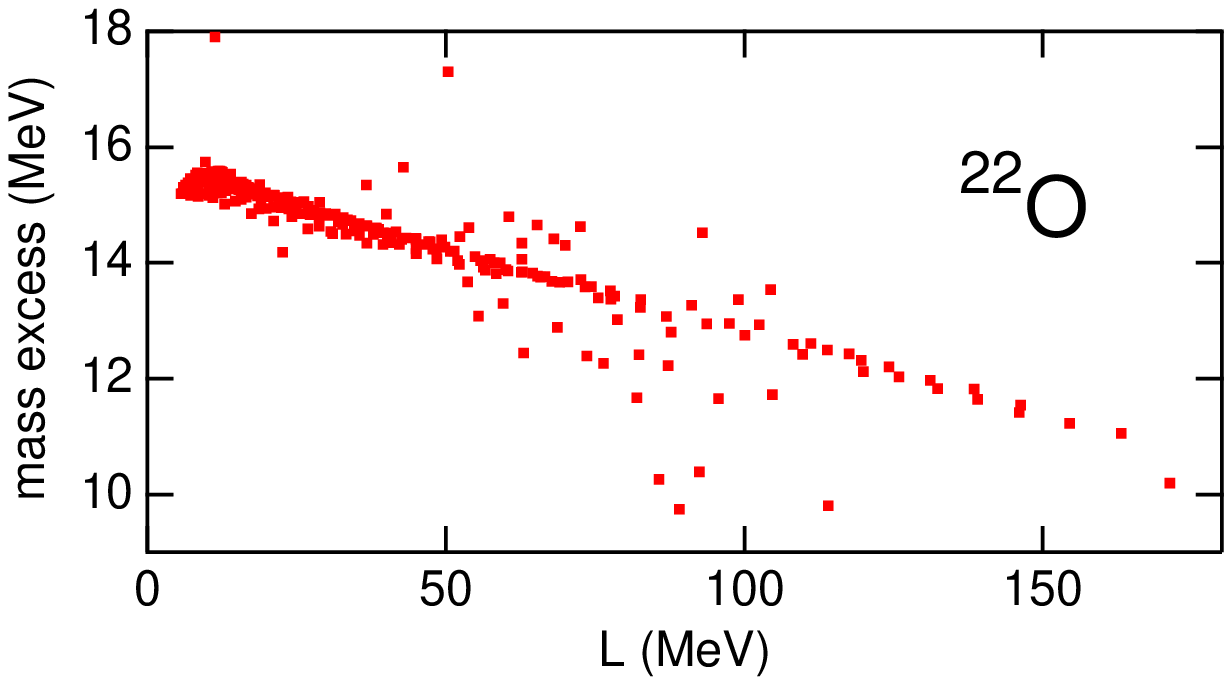}
 \end{minipage}
\vspace{-0.5cm}
\caption{\label{massNi} (Color online)
The mass excess calculated for $^{78}$Ni and $^{22}$O as a function of $L$.
}
\end{figure}

     The effect of $a_{\rm ssym}$ can be understood by considering
the density-dependent surface tension \cite{IO},
\begin{equation}
\sigma(n_{\rm in},\alpha_{\rm in})
=\sigma_0\left[1-C_{\rm sym}\alpha_{\rm in}^2
+\chi\left(\frac{n_{\rm in}-n_0}{n_0}\right)\right],
\end{equation}     
where $n_{\rm in}$ and $\alpha_{\rm in}$ are the density and neutron 
excess inside a liquid-drop, $\sigma_0=\sigma(n_0,0)$, $C_{\rm sym}$ is 
the surface symmetry energy coefficient, and $\chi=(n_0/\sigma_0)
\partial\sigma/\partial n_{\rm in}|_{n_{\rm in}=n_0, \alpha_{\rm in}=0}$.
By substituting $n_s$ into $n_{\rm in}$, one obtains
\begin{equation}
\sigma(n_s,\alpha_{\rm in})
=\sigma_0\left[1-\left(C_{\rm sym}+\frac{3L\chi}{K_0}\right)
\alpha_{\rm in}^2\right].
\label{ndsigma}
\end{equation}     
Thus, $a_{\rm ssym}A^{2/3}$ behaves as 
$-4\pi \sigma_0 R^2 \left(C_{\rm sym}+\frac{3L\chi}{K_0}\right)$, with
the liquid-drop radius $R$.   The condition $\chi>0$, which is suggested by
various nuclear models including the macroscopic one utilized here 
\cite{note}, is desirable for understanding of the $L$ dependence of 
$\delta V_{np}$.  We remark that this condition is suggested by various 
mean-field models because the calculated neutron skin thickness, which is 
basically proportional to $-a_{\rm ssym}$, increases with $L$ \cite{Warda}.

\section{Conclusions}
\label{sec:con}

      We have analyzed the influence of the density dependence of
the symmetry energy on nuclear masses by using a macroscopic nuclear model
that depends explicitly on the EOS of nuclear matter.  We find that
the $L$ dependence of the calculated masses comes mainly from the surface
property through the density and neutron excess dependence of the
surface tension.

       In making reasonable estimates of $L$ from empirical masses, we 
find that empirical two-proton separation energies for neutron-rich, light 
nuclei are useful, implying a larger $L$ value.  However, we still have two 
caveats.   First, Eq.\ (\ref{ndsigma}) suggests that uncertainties in 
$\chi/K_0$ affect the prediction of the value $L$.  Second, a smooth isotope 
dependence of empirical mass data is hard to derive.  Consequently, such 
estimates of $L$ are closely connected to determination of the parameters
$\chi$ and $K_0$ characterizing the density dependence of the surface and
bulk energy and to understanding of the discrete behavior of $S_{2p}$ and 
$\delta V_{np}$ \cite{Stoitsov,Chen,GSKH}.

\acknowledgments

      We are grateful to Drs.\ H. Koura and A. Kohama for useful discussion
on this and related subjects.  We acknowledge the hospitality of the
Yukawa Institute for Theoretical Physics during the workshop ``New Frontiers
in QCD 2010,'' where this work was finished.


\begin{references}

\bibitem{BW} J.M. Blatt and V.F. Weisskopf, {\it Theoretical Nuclear Physics}
(Wiley, New York, 1952).
\bibitem{HP} H. Heiselberg and V.R. Pandharipande, Ann.\ Rev.\ Nucl.\ Part.\ 
Sci.\ {\bf 50}, 481 (2000).
\bibitem{OTSST} K. Oyamatsu, I. Tanihata, Y. Sugahara, K. Sumiyoshi, and 
H. Toki, Nucl.\ Phys.\ {\bf A634}, 3 (1998).
\bibitem{B} B.A. Brown, Phys.\ Rev.\ Lett.\ {\bf 85}, 5296 (2000).
\bibitem{CWS} K.C. Chung, C.S. Wang, and A.J. Santiago, nucl-th/0102017.
\bibitem{OI} K. Oyamatsu and K. Iida, Prog.\ Theor.\ Phys.\ {\bf 109}, 631
  (2003).
\bibitem{Daniel} P. Danielewicz, R. Lacey, and W.G. Lynch, Sci.\ {\bf 298},
1592 (2002).
\bibitem{LCK} B.A. Li, L.W. Chen, and C.M. Ko, Phys.\ Rep.\ {\bf 464}, 113
(2008).
\bibitem{YCL} D.H. Youngblood, H.L. Clark, and Y.-W. Lui, Phys.\ Rev.\ 
Lett.\ {\bf 82}, 691 (1999).
\bibitem{Pons} J.M. Lattimer and M. Prakash, Sci.\ {\bf 304}, 536 (2004).
\bibitem{Guillot} S. Guillot, R.E. Rutledge, L. Bildsten, E.F. Brown, 
G.G. Pavlov, and V.E. Zavlin, Mon.\ Not.\ R. Astron.\ Soc.\ 
{\bf 392}, 665 (2009).  
\bibitem{L} J.M. Lattimer, Ann.\ Rev.\ Nucl.\ Part.\ Sci.\ {\bf 31}, 337 
(1981).
\bibitem{IO} K. Iida and K. Oyamatsu, Phys.\ Rev.\ C {\bf 69}, 037301 (2004).
\bibitem{Warda} M. Warda, X. Vi{\~n}as, X. Roca-Maza, and M. Centelles, 
Phys.\ Rev.\ C {\bf 80}, 024316 (2009).
\bibitem{FP} B. Friedman and V.R. Pandharipande, Nucl.\ Phys.\ {\bf A361},
 502 (1981).
\bibitem{LP} I.E. Lagaris and V.R. Pandharipande, Nucl.\ Phys.\ {\bf A369},
 470 (1981).
\bibitem{O} K. Oyamatsu, Nucl. Phy.\ {\bf A561}, 431 (1993).
\bibitem{OI2} K. Oyamatsu and K. Iida, Phys.\ Rev.\ C {\bf 75}, 015801 (2007).
\bibitem{constraints} See, e.g., Fig.\ 6 in Ref.\ \cite{Warda}; G. Col{\`o},
N. Van Giai, J. Meyer, K. Bennaceur, and P. Bonche,
Phys.\ Rev.\ C {\bf 70}, 024307 (2004). 
\bibitem{isobar} A. Ozawa {\it et al.}, Phys.\ Lett.\ {\bf B344}, 18 (1994);
L. Chulkov {\it et al.}, Nucl.\ Phys.\ {\bf A603}, 219 (1996); A. Ozawa
{\it et al.}, Nucl.\ Phys.\ {\bf A709}, 60 (2002).
\bibitem{Audi} G. Audi, A.H. Wapstra, and C. Thibault,
Nucl.\ Phys.\ {\bf A729}, 337 (2003).
\bibitem{dV} H. de Vries, C.W. de Jager, and C. de Vries, At.\ Data Nucl.\ Data
Tables {\bf 36}, 495 (1987).
\bibitem{YM} M. Yamada and Z. Matumoto, J. Phys.\ Soc.\ Jpn.\ {\bf 16}, 1497 
(1961).
\bibitem{Y} M. Yamada, Prog.\ Theor.\ Phys.\ {\bf 32}, 512 (1964).
\bibitem{KTUY} H. Koura, T. Tachibana, M. Uno, and M. Yamada, 
Prog.\ Theor.\ Phys.\ {\bf 113}, 305 (2005).
\bibitem{Stoitsov} M. Stoitsov, R.B. Cakirli, R.F. Casten, W. Nazarewicz, and
W. Satu{\l}a, Phys.\ Rev.\ Lett.\ {\bf 98}, 132502 (2007).
\bibitem{note} For example, the Fermi gas model gives $\chi=4/3$.
\bibitem{Chen} L. Chen {\it et al.}, Phys.\ Rev.\ Lett.\ {\bf 102}, 
122503 (2009).
\bibitem{GSKH} A. Gelberg, H. Sakurai, M.W. Kirson, and S. Heinze, Phys.\ 
Rev.\ C {\bf 80}, 024307 (2009).
\end{references}
\end{document}